\documentclass[aps,twocolumn,amsmath,amssymb,amsfonts,floatfix,superscriptaddress]{revtex4-1}
\usepackage{dcolumn}  
\usepackage{multirow}
\usepackage[T1]{fontenc}
\usepackage{verbatim}
\usepackage{bm} 
\usepackage{hyperref}
\usepackage{natbib}
\usepackage[english]{babel}
\usepackage{babelbib}
\usepackage{graphicx}
\usepackage{epstopdf}
\usepackage{ifpdf} 
\usepackage{epsfig}
\usepackage{color}
\usepackage[usenames,dvipsnames]{xcolor}
\usepackage{mathrsfs}

\newcommand{\rmd}{{\mathrm{d}}}

\newcommand{\kBT}{k_{\mathrm{B}}T}

\begin{document}

\title{Active particles with polar alignment in ring-shaped confinement}

\author{Zahra Fazli}
\thanks{z.fazli@ipm.ir (corresponding author)}
 \affiliation{School of Physics, Institute for Research in Fundamental Sciences (IPM), Tehran 19395-5531, Iran}
\author{Ali Naji}
\thanks{a.naji@ipm.ir}
 \affiliation{School of Physics, Institute for Research in Fundamental Sciences (IPM), Tehran 19395-5531, Iran}
 \affiliation{School of Nano Science, Institute for Research in Fundamental Sciences (IPM), Tehran 19395-5531, Iran}
  
\begin{abstract}
We study steady-state properties of a suspension of  active, nonchiral and chiral,  Brownian particles with polar alignment and steric interactions confined within a ring-shaped (annulus) confinement in two dimensions. Exploring possible interplays between polar interparticle alignment, geometric confinement and the surface curvature, being incorporated here on minimal levels, we  report a surface-population reversal effect, whereby active particles migrate from the outer concave boundary of the annulus to accumulate on its inner convex boundary. This contrasts the conventional picture, implying stronger accumulation  of active particles on concave boundaries relative to the convex ones. The population reversal is caused by both particle alignment and surface curvature, disappearing when either of these factors is absent. We explore the ensuing consequences for the chirality-induced current and swim pressure of active particles and analyze possible roles of system parameters, such as the mean number density of particles and particle self-propulsion, chirality and  alignment strengths.
\end{abstract}

\maketitle

\section{Introduction}

Active fluids constitute a fascinating class of nonequilibrium systems and include numerous examples such as  bacterial suspensions and artificial nano-/microswimmers, where constituent particles, being suspended in a base fluid, display the ability to self-propel by making use of internal mechanisms and the ambient free energy \cite{Ramaswamy,Bechinger,Marchetti,Julicher,Vicsek,Needleman,Elgeti_2,Chate,Cates,Lauga,Saintillan,Bertin}. Being of mounting theoretical and experimental interest in recent years, active fluids are found to exhibit many peculiarities, not paralleled with analogous properties in thermodynamic equilibrium \cite{Bar,Li,Bialke,Chen,Mallory,Klamser}. 

While some of the key properties of active particles, such as their excessive nonequilibrium accumulation near confining boundaries  \cite{Hernandez,Spagnolie,Schaar,Nash,Elgeti_3,Jamali} or their shear-induced behavior in an imposed flow \cite{Anand,Kaya,Rusconi,Fu,Asheichyk,Shabanniya,Ezhilan_2,Nili}, can be captured within noninteracting models, their collective properties and phase behaviors are largely determined by particle interactions, including steric and hydrodynamic interactions \cite{Burkholder,Solon_3,Singh}. Phenomenological models with polar alignment interactions, which tend to align the directions of self-propulsion of neighboring active particles, have also emerged as an important class of models, describing development of long-range orientational order  in active systems, as first proposed by Vicsek et al.  \cite{Vicsek_2,Vicsek}. Coherent collective motion, pattern formation and large-scale traveling structures, such as clusters and lanes, are among the notable phenomena observed in Vicsek-type models \cite{Chate_2,Gregoire,Martin,Sese,Grossmann}. In suspensions of active particles, steric interactions, near-field hydrodynamics  and active stresses  \cite{Delfau,Yoshinaga,Lushi,Aranson} play significant roles in engendering interparticle alignment.    

Geometric confinement has also emerged as an important factor, determining various properties  of active systems, such as active flow patterns and stabilization of dense suspensions into spiral vortices \cite{Wioland,Wioland_2,Ravnik,Doostmohammadi}.   While many interesting behaviors (e.g., capillary rise of active fluids in thin tubes, shape deformation of vesicles enclosing active particles,  upstream swim in channel flow, etc; see Refs. \cite{Li,Wysocki,Kantsler} and references therein) have been reported to occur due to the near-surface behavior of active particles, geometric constraints have also been utilized in useful applications, such as steering  living or artificial microswimmers in a desired direction \cite{Das_2,Ostapenko,Popescu}. 

Active particles can lead to intriguing attractive and  repulsive effective interactions \cite{Leite,Mokhtari,Zaeifi,Zarif,Sebtosheikh} between passive colloidal objects and rigid surface boundaries as well, with interaction profiles that exhibit rather complex dependencies on system parameters, including the distance between the juxtaposed surfaces. The latter arises due partly to the aforementioned boundary accumulation of active particles  that can lead to pronounced near-surface  particle layering   \cite{Zaeifi,Zarif,Sebtosheikh,Ni,Ye}. The surface accumulation of active particles has been associated  with a number of different factors, including hydrodynamic particle-wall couplings  \cite{Elgeti,Elgeti_3,Hernandez,Li_5,Spagnolie} and, on a more basic level, the persistent motion of the particles \cite{Schaar,Wysocki_2}, causing prolonged near-surface detention times. A remarkable manifestation of the foregoing effects is the so-called swim pressure produced by active particles on confining boundaries, a notion of conceptual significance in describing the steady-state properties of active fluids via possible analogies with equilibrium thermodynamics that has attracted much  interest and debate in the recent past \cite{Takatori,Fily_2,Chen,Winkler,Solon_2,Junot,Marconi,Speck_2,Solon_3,Yan,Yan_2,Ginot,Patch,Marconi_2,Solon,Smallenburg,Ezhilan}.  Swim pressure and effective coupling between active particles and curved  boundaries are known to result in a diverse range of phenomena, including negative interfacial tension \cite{Bialke,Speck_3},  bidirectional flows and helical vortices in cylindrical capillaries  \cite{Ravnik}, microphase separation  \cite{Singh,Tung} , and reverse/anomalous Ostwald ripening \cite{Jamali,Tjhung},  
with the latter mechanism enabling stabilized  mesophases of monodispersed droplets   (see also other related works on active droplets in Refs. \cite{Zwicker,Zwicker_2,Weber,Golestanian}).

In this work, we study surface accumulation and swim pressure of active, nonchiral and chiral, Brownian particles within a minimal two-dimensional model by combining the three main ingredients noted above, i.e., the steric and polar interparticle alignment interactions, geometric confinement and boundary curvature. The two latter factors are brought in by adopting a ring-shaped (annulus) confinement. This geometry is  particularly useful in that it allows one  to examine the interplay between convex and concave curvatures introduced by its circular inner and outer boundaries, respectively. The system is described in the steady state using a Smoluchowski equation, governing the joint position-orientation  probability distribution function of active particles, which is numerically analyzed to derive quantities such as particle density and polarization profiles, chirality-induced current and swim pressure on the inner and outer circular boundaries of the annulus. We thus report a surface-population reversal, whereby  active particles accumulate more strongly near the convex inner boundary of the annulus rather than its concave outer boundary. This contrasts the conventional picture implying preferential  accumulation (due to longer detention times) of active particles near concave boundaries relative to the convex ones, a behavior that is found for noninteracting particles within the present model as well. We thus show that the said population reversal  is a direct consequence of both  alignment interactions and boundary curvature and will be absent in the absence of either of them. The implications of this effect for the chirality-induced current and swim pressure of active particles are then explored in detail. We describe our model and its governing equations in Sections   \ref{II} and \ref{III}, discuss our results in Section \ref{IV}, and conclude the paper in Section  \ref{V}. 

\begin{figure*}[t!]
\centering
\includegraphics[width=0.95\textwidth]{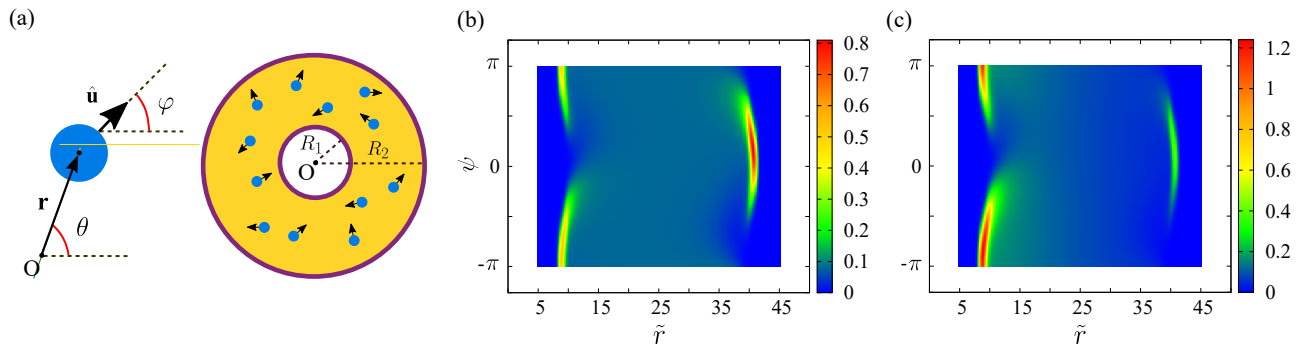}
\caption{
(a)  Left: Active particles are described by their position and orientation vectors parametrized in polar coordinates as ${\mathbf r}=r(\cos\theta, \sin\theta)$ and $\hat{\mathbf{u}}=(\cos\varphi,\sin\varphi)$; Right: They are confined within a ring-shaped confinement, or annulus, with impermeable circular boundaries of radii $R_1$ and $R_2$ from the origin O. (b) and (c) Steady-state PDF, $\tilde{\cal P}_0(\tilde r, \psi)$ of active  particles within the shown confined geometry is plotted in the $\psi-\tilde r$ plane for   nonaligning  (panel b, $U_0=0$) and interaligning particles (panel c, $U_0=10$) with fixed  parameter values are $\tilde R_1=10$, $\tilde R_2=40$, ${\rm Pe}=10$, $\tilde \rho_0=0.1$ and $\Gamma=1$. 
}
\label{fig1}
\end{figure*}

\section{Model}\label{II}

We consider a two-dimensional minimal model of active Brownian  particles   \cite{Marchetti_2}  confined between two impermeable, concentric, circular boundaries with radii $R_{1}$ and $R_{2}>R_1$, forming a ring-shaped (annulus) confinement; see Fig. \ref{fig1}a. Each particle has an intrinsic, constant, self-propulsion speed $v>0$ along its instantaneous direction of motion, determined by the unit vector $\hat{\mathbf{u}}$, and may also possess an intrinsic, constant, in-plane angular velocity (chirality) of  signed magnitude $\omega$; see, e.g., Refs. \cite{Liebchen,Ai,Mijalkov,Nourhani,Lowen}. Particles are assumed to interact via pair potentials of two main types: (i) a phenomenological alignment interaction of strength $U_0>0$ to be modeled through the dot-product of particle orientations as $-U_0\,\hat{\mathbf{u}}_i\cdot\hat{\mathbf{u}}_j$ for the pair of particles labeled by $i$ and $j$ \cite{Fazli}; (ii) a local steric delta-function potential  $U_1\,\delta({\mathbf r}_i-{\mathbf r}_j)$ for the given pair. The latter provides a formally straightforward  route to incorporate a finite excluded area $U_1$ for each particle in our later continuum formulation. Thus, the total pair interaction energy ${\mathcal U}_{ij}\equiv{\mathcal U}({\mathbf r}_i, \hat{\mathbf{u}}_i; {\mathbf r}_j, \hat{\mathbf{u}}_j)$, is
\begin{equation}
\frac{{\mathcal U}_{ij}}{\kBT }=-U_0\,\hat{\mathbf{u}}_i\cdot\hat{\mathbf{u}}_j+U_1\,\delta({\mathbf r}_i-{\mathbf r}_j).
\end{equation}

Particles interact with the circular boundaries of the confinement with the repulsive simple harmonic potential 
\begin{equation}\label{harm_pot}
\!{\mathcal V}_i\!=\!\frac{k_{1}}{2}\!\left(r_i-R_{1}\right)^2\!\Theta(R_{1}-r_i)+\frac{k_{2}}{2}\!\left(r_i-R_{2}\right)^2\!\Theta(r_i-R_{2}),
\end{equation}
where ${\mathcal V}_i\equiv {\mathcal V}(r_i)$ and $r_i$ is the radial distance of the $i$th particle from the origin, see Fig. \ref{fig1}, $\Theta(\cdot)$ is the Heaviside step function,  and $k_{1}=k_{2}$ are the respective harmonic constants. These constants are fixed at a sufficiently large value to establish  nearly hard boundaries, being also  assumed to be torque-free  (no dependence of ${\mathcal V}_i$ on the particle orientation); hence,  the swim pressure on them is expected to be a state function  \cite{Solon_2,Fily_2}. It shows no dependence on the  interfacial potential strengths.

\section{Governing equations}\label{III}

The  translational and rotational dynamics of active particles are described by the  Langevin equations 
\begin{eqnarray}\label{Lant}
&&\dot{{\mathbf r}}_i=v\hat{\mathbf{u}}_i-\mu_{\mathrm{t}}\nabla_i\left( {\mathcal V}_i+\sum_{j\neq i}{\mathcal U}_{ij}\right)+\sqrt{2D_{\mathrm{t}}}\boldsymbol\eta_i(t),\\
\label{Lanr}
&&\dot{\varphi}_i=\omega-\mu_{\mathrm{r}}\partial_\varphi\sum_{j\neq i}{\mathcal U}_{ij}+\sqrt{2D_{\mathrm{r}}}\zeta_i(t), 
\end{eqnarray}
with the shorthand notations  $\nabla_i\!=\!\partial/\partial {\mathbf r}_i$ and $\!\partial_\varphi\!=\!\partial/\partial \varphi$. Here,  ${\boldsymbol\eta}_i$ and $\zeta_i$ are the translational and rotational white noises of zero mean and unit variances, with $D_{\mathrm{t}}$ and $D_{\mathrm{r}}$ being the single-particle translational and rotational diffusivities, respectively. To ensure that the system behavior  reduces to that of an analogues equilibrium system, when the active sources are put to zero, $v=\omega=0$, the diffusivities are assumed to fulfill the Smoluchowski-Einstein-Sutherland relations $D_{\mathrm{t}}=\mu_{\mathrm{t}}\kBT$ and $D_{\mathrm{r}}=\mu_{\mathrm{r}}\kBT$, where $\kBT$ is the ambient thermal energy scale and $\mu_{\mathrm{t}}$ and $\mu_{\mathrm{r}}$ are the  translational and rotational (Stokes) mobilities, respectively \cite{Happel}.  

Equations (\ref{Lant}) and (\ref{Lanr}) can standardly be mapped to a time-evolution equation involving one- and two-point  PDFs, with the latter originating from the two-particle interactions.  Without delving into further details, we adopt a mean-field approximation as a closure scheme for such an equation by neglecting interparticle correlations and assuming  ${\cal P}({\mathbf r},\varphi, {\mathbf r}',\varphi'; t)={\cal P}({\mathbf r},\varphi; t){\cal P}({\mathbf r}',\varphi'; t)$, which is expected to hold in sufficiently dilute suspensions. This leads to an effective Smoluchowski equation, involving an effective translational flux velocity $-\mu_{\mathrm{t}}(\nabla \overline{{\mathcal U}})$ experienced on average by each particle due to its interactions with other particles, where the averaged two-particle interaction $\overline{{\mathcal U}}=\int \rmd {\mathbf r}'\rmd \varphi'\,{\mathcal U}\,{\cal P} ({\mathbf r}',\varphi')$. Such an equation is nonlocal and nonlinear and can be expressed as 
\begin{eqnarray}\label{eq:SE}
&&\partial_t{\cal P}=-\nabla\cdot\left[\left(v\hat{\mathbf{u}}-\mu_{\mathrm{t}}\nabla ({\mathcal V}+\overline{{\mathcal U}})\right){\cal P}-D_{\mathrm{t}}\nabla{\cal P}\right]\nonumber\\
&&\qquad\quad -\partial_\varphi\left[\left(\omega-\mu_{\mathrm{r}}\partial_\varphi\, \overline{{\mathcal U}}\right){\cal P}-D_{\mathrm{r}}\partial_\varphi{\cal P}\right]. 
\end{eqnarray}
Our focus will be on the steady-state solution $\tilde{{\cal P}}_0({\mathbf r},\varphi)$. Because of the rotational symmetry in the present geometry, the angular dependence of the PDF can be assumed to occur only through the relative angle $\psi=\varphi-\theta$, allowing the relations  $\partial_\varphi{\cal P}=-\partial_\theta{\cal P}=\partial_\psi{\cal P}$ and $\int\rmd {\mathbf r}'\rmd \varphi'\equiv\int r' \rmd r'\rmd \psi'$ to be used, where necessary. In this case, the mean-field interaction energy reads   
\begin{equation}\label{mean_int}
\frac{\overline{{\mathcal U}}}{\kBT }\!=\!-U_0\!\!\int\!\! r'\! \rmd r'\!\rmd \psi'\!\cos(\psi'\!-\psi){\cal P}_0 (r'\!,\psi')+U_1\!\!\int\!\!\rmd \psi'{\cal P}_0(r,\psi').
\end{equation}
 Equation \eqref{eq:SE} is supplemented by  no-flux boundary conditions in the radial direction on the two circular boundaries at $r=R_1$ and $R_2$ and periodic boundary condition over the angular coordinate $\psi$ for $r\in [R_1, R_2]$. Since we have assumed a closed system with a fixed number $N$ of active particles within the annulus, the normalization condition for the PDF can be expressed using the  mean number density $\rho_0=N/[\pi({R}_2^2-{R}_1^2)]$ of particles  as 
\begin{equation}
\label{eq:norm}
\int_{ R_1}^{ R_2}\!\int_{-\pi}^{\pi} r\,\rmd r \,\rmd\psi \,{\cal P}_0( r,\psi)=\pi\rho_0\left(R_2^2-R_1^2\right). 
\end{equation}

\begin{figure*}[t!]
\centering
\includegraphics[width=0.85\textwidth]{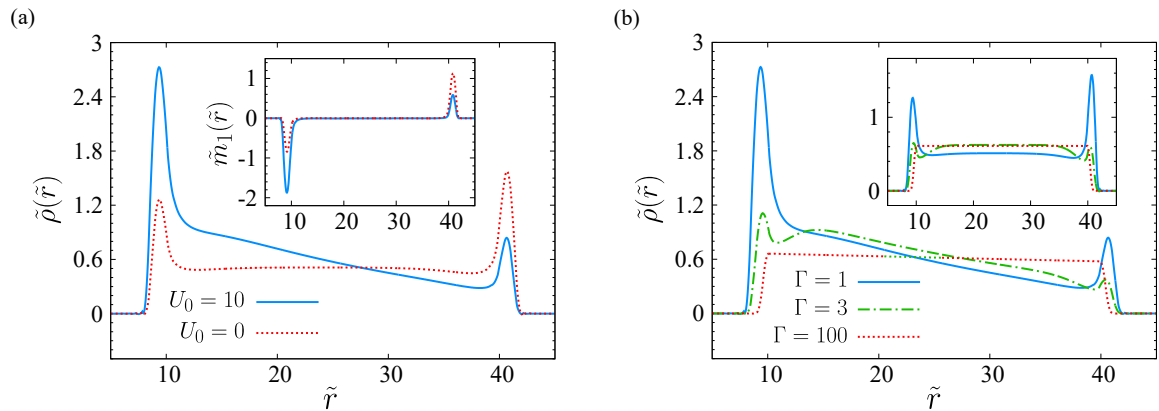}
\caption{
 (a) Rescaled number density profile of active particles, $\tilde \rho(\tilde r)$, as a function of the radial distance inside the annulus in the absence ($U_0=0$) and presence ($U_0=10$) of alignment interactions. Inset shows the corresponding polarization profiles, $\tilde m_1(\tilde r)$. Here, we have fixed $\Gamma=1$ and other parameters are as in Fig. \ref{fig1}. (b) Same as (a) but plotted for different values of chirality strength, as indicated on the plot both without  (inset, $U_0=0$) and with (main set, $U_0=10$) alignment interactions. 
}
\label{fig2}
\end{figure*}

\subsection{Dimensionless representation}

We proceed by rescaling units of length/time with the characteristic length/time $a=({D_{\mathrm{t}}}/{D_{\mathrm{r}}})^{{1}/{2}}$ 
and  $1/D_{\mathrm{r}}$, respectively, and the units of energy with $\kBT$; thus, e.g., the boundary interaction energy is rescaled as  $\tilde{{\mathcal V}}(\tilde r)={\mathcal V}(a \tilde r)/({\kBT})$. The PDF is suitably rescaled as  $\tilde{{\cal P}}_0(\tilde{r}, \psi)={\cal P}_0(a\tilde{r}, \psi)/\rho_0$. The parameter space is thus spanned by the set of dimensionless parameters defined by the {\em (swim) P{\'e}clet number} and the {\em chirality strength}  
\begin{equation}\label{peclet}
 {\rm Pe}=\frac{v}{aD_{\mathrm{r}}},\quad\Gamma=\frac{\omega}{D_{\mathrm{r}}}, 
\end{equation}
respectively, the rescaled interaction parameters  $U_0$, $\tilde{U}_1={U_1}/{a^2}$, and $\tilde{k}_{1,2}={k_{1,2} a^2}/({\kBT })$, the rescaled radii $\tilde{R}_{1,2}={R_{1,2}}/{a}$, and the rescaled mean particle density $ \tilde \rho_0 = \rho_0a^2$.  The steady-state Smoluchowski equation can then be expressed in dimensionless units as 
\begin{equation}\label{gov}
\frac{1}{\tilde{r}}\partial_{\tilde{r}}\left(\tilde{r}\tilde{{\cal J}}_{\tilde{r}}\right)+\frac{1}{\tilde{r}}\partial_{\psi}\tilde{{\cal J}}_\psi=0,
\end{equation}
where we have defined the rescaled spatial and angular probability flux densities $\tilde{{\cal J}}_{\tilde{r}}$ and $\tilde{{\cal J}}_\psi $, respectively, as  
\begin{widetext}
\begin{eqnarray}\label{j1_s}
&&\tilde{{\cal J}}_{\tilde{r}}=-\partial_{\tilde{r}}\tilde{{\cal P}}_0+\left({\rm Pe}\cos\psi-\partial_{\tilde{r}}\tilde{{\mathcal V}}\right)\tilde{{\cal P}}_0-\tilde{U}_1\tilde{\rho}_0\left(\partial_{\tilde{r}}\!\!\int\!\rmd \psi'\tilde{{\cal P}}_0(\tilde{r},\psi')\right)\tilde{{\cal P}}_0,\\
\label{j2_s}
&&\tilde{{\cal J}}_\psi =-\frac{1}{\tilde{r}}\left(1+\tilde{r}^2\right)\partial_\psi\tilde{{\cal P}}_0+\left(-{\rm Pe}\sin\psi+\Gamma\tilde{r}\right)\tilde{{\cal P}}_0+\frac{U_0\tilde{\rho}_0}{\tilde{r}}\left(1+\tilde{r}^2\right)\left(\int{\tilde{r}}'\rmd {\tilde{r}}'\rmd \psi'\sin(\psi'-\psi)\tilde{{\cal P}}_0(\tilde{r}',\psi')\right)\tilde{{\cal P}}_0, 
\end{eqnarray}
\end{widetext}
where 
$\int_{\tilde R_1}^{\tilde R_2}\int_{-\pi}^{\pi}\tilde r\rmd\tilde r \rmd \psi \tilde{\cal P}_0(\tilde r,\psi)=1$. Equation \eqref{gov} is an integrodifferential PDE, which is solved using finite-element methods within the annulus subject to the aforementioned boundary conditions. Other rescaled steady-state quantities such as the particle density profiles within the annulus and swim pressure on the boundaries follow directly from the numerically obtained PDF \cite{Jamali}. The choice of dimensionless parameter values and their relevance to realistic systems is discussed in Appendix \ref{app:parameters}.  Since Eq. \eqref{gov} remains invariant under chirality and orientation angle reversal, $\Gamma\rightarrow -\Gamma$ and $\psi\rightarrow -\psi$, we restrict our discussion to only positive (counterclockwise) values of $\Gamma$, bearing in mind that chirality-induced currents will be reversed for negative values of $\Gamma$.   

\section{Results}\label{IV}

\subsection{Particle distribution and radial density profile} 
\label{subsec:PDF}

In Figs. \ref{fig1}b and c, the steady-state PDF of active particles inside the annulus is plotted in the $\psi-\tilde r$ plane in the absence (panel b, $U_0=0$) and  presence (panel c, $U_0=10$) of alignment  interactions between the particles  for a representative set of parameter values. As seen, most of active particles accumulate near the boundaries (represented by larger probability densities appearing in yellow and red colors in the plots),  a well-known consequence of the persistent motion of particles \cite{Elgeti}. Also, in both the absence and the presence of alignment  interactions,  the typical orientations of particles on the outer (inner) boundaries are expectedly in the outward (inward)  directions pointing away (toward) the origin of the annulus, respectively. The PDF of nonchiral active particles  \cite{Jamali} would be peaked exactly at $\psi=0$ and $\psi=\pm\pi$ on the outer  and inner boundaries, respectively, reflecting the normal-to-boundary orientations of the particles. In the plots, however, we have taken a  relatively small, counterclockwise, particle chirality ($\Gamma=1$) that produces a shift of the  probability-density peaks from the noted angular orientations to the first and third angular quadrants near the outer and inner boundaries, respectively. 

Also, as seen in Fig. \ref{fig1}b, nonaligning particles are more strongly accumulated near the outer boundary than the inner one, in accord with the standard paradigm that active particles spend longer `detention' times at concave rather  than convex boundaries \cite{Fily_3,Nikola,Paoluzzi}. Remarkably, we find a reversed situation in the presence of interparticle alignment (panel c) with larger particle probabilities appearing near the inner boundary. This appears to suggest an alignment-induced crossover between two different configurational states with active particles  primarily attracted to only one of the boundaries. Such a behavior, which shall examine more closely later, can also be discerned from the local density of active particles inside the annulus defined as 
\begin{equation}
\tilde\rho(\tilde r)=\int_{-\pi}^{\pi}\rmd \psi \, \tilde {\cal P}_0(\tilde r,\psi). 
\end{equation}
This quantity being plotted in Fig.  \ref{fig2}a indicates a reversal in the relative accumulation of active particles near the two boundaries, when the alignment interaction strength $U_0$ is changed from $U_0=0$ (dotted curve) to $U_0=10$ (solid curve). The figure also shows that the wide plateau present in the density profile of nonaligning particles changes to a linear stretch of particle density between the inner and outer peaks, when the alignment interaction is switched on. The active particles within this linear region of the density profiles turn out to show no specific orientational order, as it can be verified using the first angular moment of the PDF defined as 
\begin{equation}
\tilde m_1(\tilde r)=\int_{-\pi}^{\pi}\rmd \psi\cos(\psi) \tilde{\cal P}_0(\tilde r,\psi). 
\label{eq:m1}
\end{equation}
Being shown in the inset of Fig. \ref{fig2}a, $\tilde m_1(\tilde r)$  reflects the expected outward- (inward-) pointing mean polarization of the particles near the outer (inner) boundaries but no mean orientational order ($\tilde m_1\simeq 0$) elsewhere across the central regions, neither for nonaligning nor for interaligning particles. 

When the chirality strength $\Gamma$, is increased, as shown in Fig. \ref{fig2}b, the near-boundary accumulation of active particles is suppressed in both the absence (inset, $U_0=0$) and the presence (main set, $U_0=10$) of alignment  interactions. One  eventually finds  a homogeneous density profile at elevated  $\Gamma$  (see the red dotted curve for $\Gamma=100$), consistent with known results \cite{Ao,Li_2}, indicating that the behavior of chiral active particles reduces to that of passive particles in the limit of infinite chirality (see Ref. \cite{Jamali} for a  systematic derivation). At intermediate  $\Gamma$, however, our results reveal nontrivial variations in the density profile, with the density peak at the inner (outer) boundaries  followed (preceded) by a shallow dip, indicating partial depletion of active particles (green dot-dashed curve for  $\Gamma=3$). This is a consequence of the fact that chirality effects in suppressing the persistent motion of active particles are more apparent near the boundaries and the density peaks are suppressed  more strongly than the linear bridge connecting them, as $\Gamma$ is increased.

\begin{figure}[t!]
\centering
\includegraphics[width=0.85\linewidth]{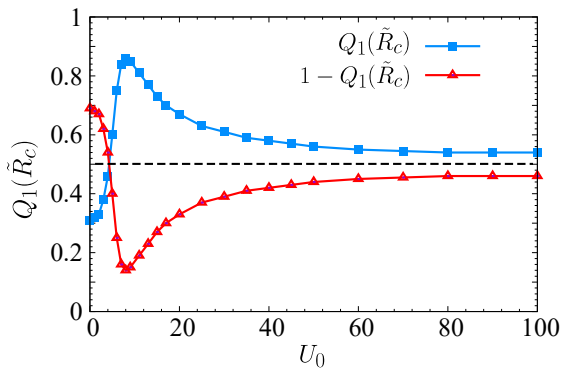}
\caption{Mean fraction of active particles found within the inner and outer semi-annuli, $Q_1(\tilde{R}_c)$ and $1-Q_1(\tilde{R}_c)$, respectively (see the text for definitions), as functions of the alignment interaction strength for fixed parameter values $\tilde{R}_1=10$, $\tilde{R}_2=40$, ${\rm Pe}=10$, $\tilde{\rho}_0=0.01$, $\Gamma=1$.  Symbols are numerical data and curves are guides to the eye.
}
\label{fig3}
\end{figure}

\begin{figure}[t!]
\centering
\includegraphics[width=0.83\linewidth]{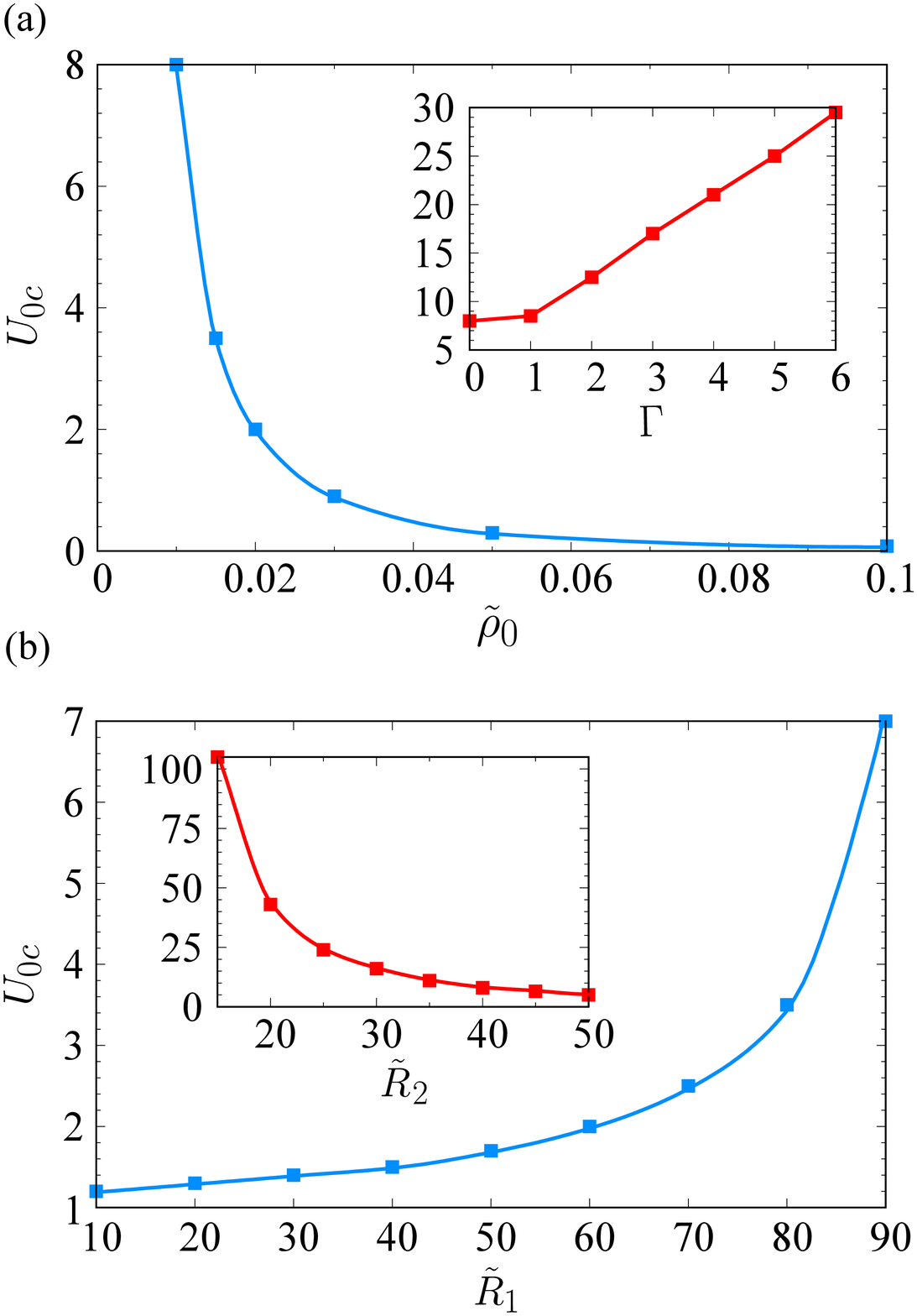}
\caption{(a) Dependence of $U_{0c}$, the alignment strength giving the maximum population reversal,  on the mean number density (main set) and chirality strength (inset) of active particles for fixed  $\tilde{R}_1=10$, $\tilde{R}_2=40$, ${\rm Pe}=10$, and with $\Gamma=1$ (main set) and $\tilde{\rho}_0=0.01$ (inset). (b) Dependence of $U_{0c}$ on the radii of the inner/outer boundaries of the annulus for fixed   $\tilde{\rho}_0=0.01$, ${\rm Pe}=10$, $\Gamma=1$, and with $\tilde{R}_2=100$ (main set) and  $\tilde{R}_1=10$ (inset). Symbols are numerical data and curves are guides to the eye.  }
\label{fig4}
\end{figure}

\subsection{Surface-population reversal} 
\label{subsec:Q}

Further insight into the alignment-induced effects can be obtained by conventionally defining the inner and outer fractions $Q_1(\tilde R_c)$ and  $1-Q_1(\tilde R_c)$ of active particles as those found in radial distances $\tilde r<\tilde R_c$ and $\tilde r>\tilde R_c$, respectively, where $\tilde{R}_c=(\tilde R_1+\tilde R_2)/2$ is  the mean radius of the annulus. We thus have 
\begin{equation}
Q_1( \tilde{R}_c)=\frac{\int_{\tilde{R}_1}^{\tilde{R}_c}\,\tilde{r}\rmd \tilde{r}\tilde{\rho}(\tilde{r})}{\int_{\tilde{R}_1}^{\tilde{R}_2}\,\tilde{r}\rmd \tilde{r}\tilde{\rho}(\tilde{r})}.
\label{eq:Q}
\end{equation}
These fractions are shown as functions of the alignment interaction strength in Fig. \ref{fig3}. In the nonaligning case ($U_0=0$), a larger fraction of particles is found in the outer semi-annulus, which, as noted before, is because  self-propelling particles accumulate more strongly near concave than convex boundaries due to their lingered detention times. As seen in the figure, the inner fraction $Q_1(\tilde R_c)$ increases with $U_0$ and  $1-Q_1(\tilde R_c)$ decreases with it until they are equalized at a certain value of $U_{0\ast}$ (here, $U_{0\ast}\simeq 5$). This particular value corresponds to the onset of a counterintuitive {\em surface-population reversal}, beyond which active particles are more strongly accumulated by the {\em convex} inner boundary with the {\em smaller} radius of curvature  rather than the concave outer boundary with the larger radius of curvature. The maximum population reversal is achieved at a slightly larger value of $U_{0c}$ (here, $U_{0c}\simeq 8$), where $Q_1(\tilde R_c)$ displays a global maximum and $1-Q_1(\tilde R_c)$ a global minimum. Beyond this point, both quantities level off and gradually  tend toward 1/2, indicating an even distribution of particles developing in the infinite $U_0$ limit within the annulus (not shown). 
The designated values $U_{0\ast}$ and $U_{0c}$ are typically found to be close and, to examine the dependence of the population reversal phenomenon on other system parameters, we concentrate on the latter quantity. 

Figure \ref{fig4}a (main set) shows the dependence of the maximum population reversal plotted as a function of the mean number density of particles  inside the annulus. As seen, a larger (smaller) mean density, $\tilde{\rho}_0$, necessitates a weaker (stronger) alignment interaction strength, $U_{0c}$, to achieve the maximum population reversal.  The monotonically decreasing trend fits closely with a functional dependence of the form  $U_{0c}\sim \tilde{\rho}_0^{-\alpha}$ with  $\alpha\simeq 2$, a remarkable scaling-like behavior that remains to be understood. Figure \ref{fig4}a (inset) shows that $U_{0c}$  increases almost linearly with chirality strength, which is plausible as, for larger chirality strengths, particle directions change  more rapidly, requiring a larger value of alignment interaction strength to establish surface-population reversal.

Figure \ref{fig4}b, on the other hand,  shows that population reversal is more easily established as the difference between the radii of the inner and outer boundaries increases. At a fixed value of the outer boundary radius (here $R_2=100$, main set) or at a fixed value of the inner boundary radius (here $R_1=10$, inset), $U_{0c}$ increases as the radial width of the annulus is reduced and vice versa. This is indicative of the fact that in a narrower ring-shaped confinement, active particles tend distribute more evenly within the confinement and easily interchange between the two circular boundaries, necessitating a stronger alignment interaction to produce population reversal. 
Our numerical results (not shown) indicate that $U_{0c}$ does not significantly vary with ${\rm Pe}>1$ for which self-propulsion dominates particle diffusion (for ${\rm Pe}<1$, particles are mostly dispersed nearly uniformly  within the confinement and varying the alignment strength does not effectively change particle distribution).   

\subsection{Chirality-induced current}
\label{chiral_point}

Chiral active particles can generate net rotational currents near circular boundaries. This effect emerges as a result of the deviation of the most probable orientation of chiral particles from the normal-to-surface direction (see Section \ref{subsec:PDF}), creating   a tangential velocity component on the surface. To examine the steady-state chirality-induced current in the present context with alignment and steric interactions between particles (see Ref. \cite{Jamali} for the special case of noninteracting active particles), we begin by integrating the Smoluchowski equation over the rotational degree of freedom, $\varphi$, which  gives a relation as $\nabla\cdot\tilde{\mathbf{J}}=0$, where the current density, $\tilde{\mathbf{J}}=({\tilde{J}}_{\tilde{r}}, {\tilde{J}}_{\theta})$, has the following two components
\begin{equation}
{\tilde{J}}_{\tilde{r}}=\int_{-\pi}^{\pi}\rmd \psi\,{\tilde{A}}_{\tilde{r}}(\tilde{r},\psi),\,\,\, {\tilde{J}}_{\theta}=\int_{-\pi}^{\pi}\rmd \psi\,{\tilde{A}}_{\theta}(\tilde{r},\psi),
\end{equation}
where 
\begin{equation}
\tilde{A}_{\tilde {r}}=\left[{\rm Pe}\cos\psi-\partial_{\tilde{r}}\tilde{{\mathcal V}}\right]\tilde{{\cal P}}_0
-\partial_{\tilde{r}}\tilde{{\cal P}}_0-\tilde{U}_1\tilde{\rho}_0\tilde{{\cal P}}_0\,\partial_{\tilde{r}}\int\rmd \psi'\tilde{{\cal P}}_0,
\end{equation}
\begin{eqnarray}
\tilde{A}_{\theta}&=&{\rm Pe}\sin\psi\tilde{{\cal P}}_0-\frac{1}{\tilde{r}}\partial_{\psi}\tilde{{\cal P}}_0
\\
&+&\frac{U_0\tilde{\rho}_0}{\tilde{r}}\tilde{{\cal P}}_0\int \tilde{r}'\rmd \tilde{r}'\rmd \psi'\,\tilde{{\cal P}}_0\sin(\psi-\psi').
\nonumber
\end{eqnarray}

\begin{figure}[t!]
\centering
\includegraphics[width=0.85\linewidth]{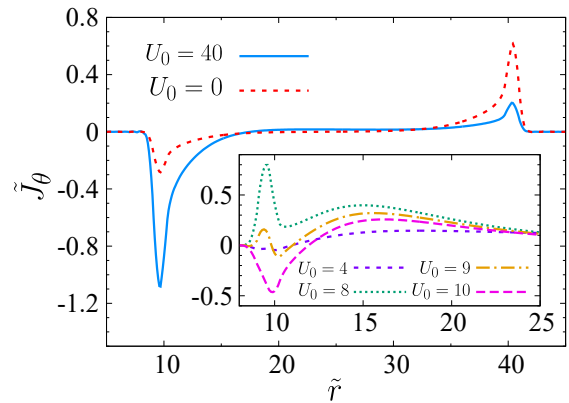}
\caption{Main set: Rescaled chirality-induced current, $\tilde{J}_\theta$, as a function of radial distance for  active particles with ($U_0=40$) and without ($U_0=0$) alignment interactions. Inset:  $\tilde{J}_\theta$ for a selected set of $U_0$ values (here, $U_{0c}=8$). Other parameters fixed as $\tilde{R}_1=10$, $\tilde{R}_2=40$, ${\rm Pe}=10$, $\tilde{\rho}_0=0.01$ and $\Gamma=1$. 
}
\label{fig5}
\end{figure}

\begin{figure*}[t!]
\centering
\includegraphics[width=0.85\textwidth]{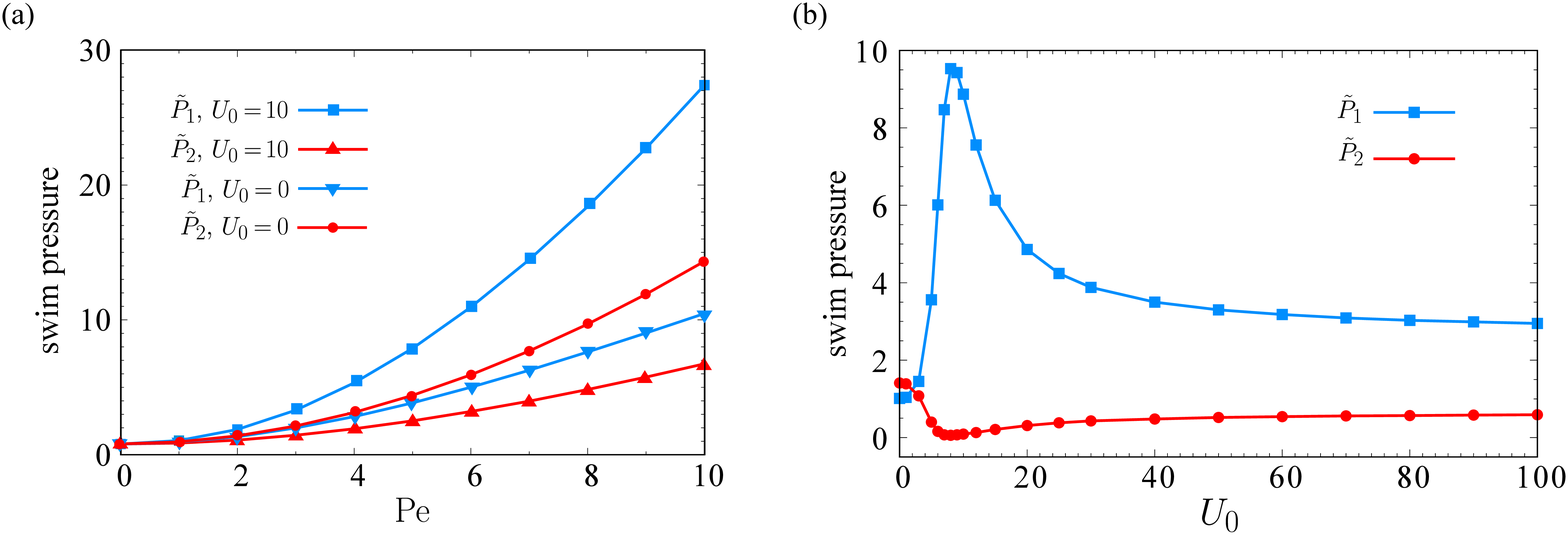}
\caption{(a) Rescaled swim pressures, $\tilde{P}_1$ and $\tilde{P}_2$ , on the inner and outer boundaries, respectively, as functions of ${\rm Pe}$ with ($U_0=10$) and without ($U_0=0$) alignment interactions for fixed  $\tilde{R}_1=10$, $\tilde{R}_2=40$, $\tilde{\rho}_0=0.1$ and $\Gamma=1$. (b) Same as (a) but plotted for swim pressures as functions of alignment interaction strength, $U_0$, for fixed $\tilde{R}_1=10$, $\tilde{R}_2=40$, ${\rm Pe}=10$, $\tilde{\rho}_0=0.01$ and $\Gamma=1$. Symbols are numerical data and curves are guides to the eye. }
\label{fig6}
\end{figure*}

The radial current density component, $\tilde{J}_{\tilde{r}}$, is zero for all parameter values. 
The angular component, $\tilde{J}_\theta$, represents the chirality-induced current and is always nonzero on the boundaries, see  Fig. \ref{fig5}, while it falls off to zero as one moves away from the boundaries, where the net current density of particles passing through a particular point vanishes \cite{Jamali}.  Figure \ref{fig5} (main set) also shows this quantity for different values of the alignment interaction strength for a representative set of parameter values  for which the maximum population reversal occurs at $U_{0c}=8$  (being nearly equal to its onset).The results represented in Fig. \ref{fig5} are typical behaviors obtained for the chirality strength $\Gamma=1$ that are shown for the sake of illustration and we can observe similar results for other representative values of $\Gamma$ near their corresponding population reversal.  Thus, while in the absence of alignment interactions ($U_0=0$, red dashed curve), the inner population shows clockwise (negative current) of smaller magnitude relative to the outer population that shows counterclockwise (positive current) of larger magnitude, the situation is reversed for active particles with strong alignment interactions ($U_0=40$, blue solid curve). 

For the rotational current on the outer boundary, increasing $U_0$ only results in a smaller current, one that never changes sign on this boundary as it diminishes with strengthening the alignment interaction. For the current on the inner boundary, we find a more complex behavior. In the inset of Fig. \ref{fig5} we show $\tilde{J}_\theta$ near the inner boundary for a few different values of alignment strength chosen around the onset of population reversal. As seen, by   increasing $U_0$ from $U_0=4$ (purple dashed curve) to $U_0=U_{0c}=8$ (green dotted curve), the shallow dip with negative rotational current close to the inner boundary turns to a region of positive current with a nonmonotonic profile with a pronounced peak. On increasing $U_0$ further to $U_0=9$ (orange dot-dashed curve) and then $U_0=10$ (pink dashed curve), the positive current profile is suppressed, even though it can now exhibit regions with both positive and negative currents ($U_0=9$) and a pronounced dip with a strong negative current ($U_0=10$) near the inner boundary. On increasing $U_0$ further, one only finds a negative current on this boundary, whose magnitude is further enhanced.

The generation of rotational current profile of varying sign near the population reversal are reminiscent of particle layering that develops at the onset of flocking transition in Vicsek-type models   \cite{Gregoire,Caussin}, even though  the ring-shaped confinement is expected to have a dominant role in the present context.

\begin{figure*}[t!]
\centering
\includegraphics[width=0.85\textwidth]{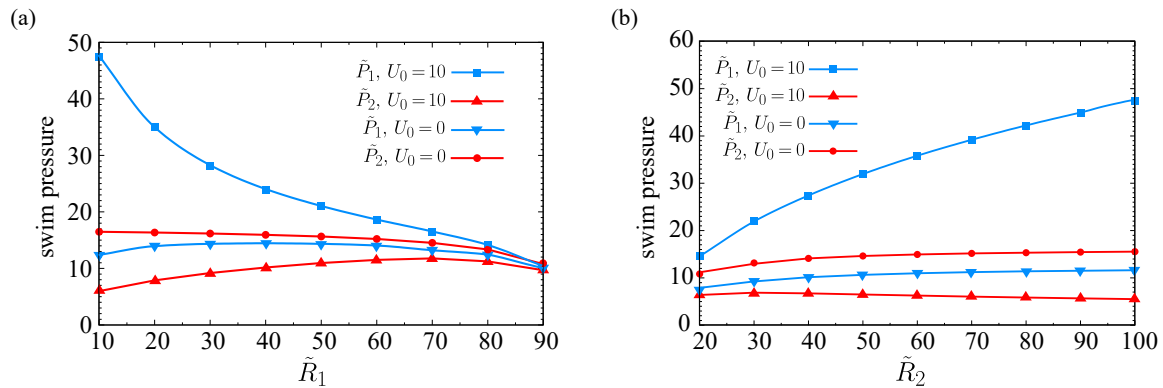}
\caption{(a) Rescaled swim pressures, $\tilde{P}_1$ and $\tilde{P}_2$, on the inner and outer boundaries, respectively, as functions of the inner radius, $\tilde{R}_1$, with ($U_0=10$) and without ($U_0=0$) alignment interactions for fixed $\tilde{R}_2=100$.  (b) Same as (a) but plotted for swim pressures as functions of the outer radius, $\tilde{R}_2$, for fixed $\tilde{R}_1=10$. Other parameters are fixed in the plots as ${\rm Pe}=10$, $\tilde{\rho}_0=0.1$ and $\Gamma=1$. Symbols are numerical data and curves are guides to the eye.  
}
\label{fig7}
\end{figure*}

\subsection{Swim pressure}\label{pressure}

Other interesting quantities we can explore in this system are swim pressures on the inner and outer boundaries, denoted by $\tilde{P}_1$ and $\tilde{P}_2$, respectively, which can be calculated by integrating the force density exerted by active particles on the boundaries as 
\begin{equation}\label{pre}
\tilde{P}_1=\int_\Lambda^0\rmd \tilde{r}\,\tilde{\rho}(\tilde{r})\partial_{\tilde{r}}\tilde{{\mathcal V}},\quad 
\tilde{P}_2=\int_\Lambda^\infty\rmd \tilde{r}\,\tilde{\rho}(\tilde{r})\partial_{\tilde{r}}\tilde{{\mathcal V}}.
\end{equation}
where  $\Lambda$ is a radial distance away from the two boundaries and inside the annulus (where $\partial_{\tilde{r}}\tilde{{\mathcal V}}$ vanishes), which we arbitrarily set equal to the mean radius of the annulus.  

The swim pressures $\tilde{P}_1$ and $\tilde{P}_2$ both turn out to be monotonically increasing functions of P\'eclet number in the presence and absence of alignment interactions, as seen in Fig. \ref{fig6}a (for the given set of fixed parameter values in the figure, the maximum population reversal occurs around $U_{0c}\simeq 0.08$; hence, the two cases shown occur are far from the onset of the reversal). For the case with aligning particles, $\tilde{P}_1$ is larger than its corresponding value in the nonaligning case (compare blue squares and triangle-downs), but $\tilde{P}_2$ shows the reverse property (compare red circles and triangle-ups). Thus, alignment interactions increase the swim pressure on the inner boundary and decrease it on the outer one. At small $\rm Pe$, $\tilde{P}_1$ and $\tilde{P}_2$ converge and tend to zero, as expected.  $\tilde{P}_1$ and $\tilde{P}_2$  diverge  as soon as $\rm Pe$ increases beyond $\rm Pe=1$. Also, for all nonvanishing $\rm Pe$, the difference between $\tilde{P}_1$ and $\tilde{P}_2$ is larger in the aligning case, which is a direct consequence of population reversal and particles departing from outer to the inner boundary.

Figure \ref{fig6}b shows $\tilde{P}_1$ and $\tilde{P}_2$ as functions of the alignment strength (for the given set of fixed parameter values, $U_{0c}\simeq 8$). As seen, $\tilde{P}_1$ first increases rapidly to a maximum value and then smoothly drops to finite values at large $U_0$. $\tilde{P}_2$ behaves in an opposite way, as it first drops to almost zero and then slowly approaches a small and finite value at large $U_0$. For $U_0=0$, the swim pressure on the inner boundary is slightly smaller than the outer one. Upon switching on the alignment interaction, particles start to migrate to the inner boundary and this causes $\tilde{P}_1$ to rapidly increase and $\tilde{P}_2$ to decrease with $U_0$ until they reach their maximum and minimum, respectively, at $U_{0c}$, with the overall trends occurring in accord with Fig. \ref{fig3}. In Appendix \ref{app1}, we derive an analytical expression for the swim pressure in the presence of alignment interactions, corroborating the foregoing discussions. 

In Fig. \ref{fig7}, we show $\tilde{P}_1$ and $\tilde{P}_2$ as functions of radii of the circular boundaries (for all values of $\tilde R_1$ and $\tilde R_2$ shown in the figure, $U_0=10$ is larger than $U_{0c}$).
In the absence of alignment interactions, $\tilde{P}_1$ and $\tilde{P}_2$ vary weakly with both $\tilde{R}_1$ and $\tilde{R}_2$ and the difference between $\tilde{P}_1$ and $\tilde{P}_2$ is small. In the presence of alignment interactions, we find a different behavior. Figure \ref{fig7}a shows that the swim pressure on the inner boundary (blue squares) decreases monotonically as a  function of $\tilde{R}_1$, attaining its maximum value at the smallest inner radii $\tilde{R}_1=10$, corresponding to the largest confinement, given that $\tilde R_2=100$ is fixed. This is reflective of stronger active-particle accumulation on the inner boundary as the system is deep inside the population-reversal regime. In this case, the swim pressure on the outer boundary  (red triangle-ups)  varies only weakly with $\tilde{R}_1$. The aforementioned behaviors are corroborated by those in Fig. \ref{fig7}b, depicting $\tilde{P}_1$ and $\tilde{P}_2$ as functions of $\tilde{R}_2$. Also, as seen in Fig. \ref{fig7}a,  all four curves are found to converge to a common value as the ring-shaped confinement becomes narrower, with $\tilde{R}_1$ tending  toward $\tilde{R}_2$. This is due to the fact  that in extremely narrow annuli, active particles tend to distribute almost evenly within the confinement with equal probabilities of interacting with either   of the boundaries, regardless of their alignment interactions. 

\section{Summary}\label{V}

In this paper, we study a two-dimensional system of nonchiral and chiral active Brownian particles constrained to move in a circular ring-shaped confinement (annulus) with impermeable confining boundaries. The active particles are assumed to have  constant intrinsic,  linear (self-propulsion) and angular, velocities and interact through alignment as well as steric pair  potentials.  The alignment interaction between particles is assumed to have a dot-product form in a way that it tends to align the self-propulsion directions of a particle pair. The steady-state properties of active particles are analyzed using a probabilistic Smoluchowski equation, which is solved numerically. This equation takes the form of an integrodifferential equation due to nonlocal coupling terms generated by the particle  interactions that are then treated using a mean-field approximation. 

While active particles are typically known to accumulate more strongly at concave rather than convex boundaries due to their longer near-boundary detention times in the former case \cite{Fily_3,Nikola,Paoluzzi,Jamali}, we show that the presence of alignment interactions causes a reverse phenomena to take place in the present setting; hence, a larger fraction of active particles are found to accumulate at the inner boundary of the annulus, which is  convex and has a smaller radius of curvature. Such a {\em surface-population reversal} is both an alignment-induced and a curvature-induced effect and will be absent in the absence of alignment interactions and/or in a planar confinement. The effect is quite robust in the sense that it emerges over a wide range of moderately large alignment interaction strengths, $U_0$,  and moderately large P\'eclet numbers (with the latter required to be only so large as to facilitate surface accumulation of active particles against particle diffusion into the bulk). The population reversal is typically maximized at an alignment strength $U_{0c}$ that is only slightly larger than the onset of the reversal, indicating a rapid crossover  at the onset, followed by a nonmonotonic behavior, i.e., a relatively sharp hump and then decay of the  surface populations down to certain saturation values, as $U_0$ is increased.  We also find that the wider the annulus (the larger the difference between the outer and inner radii) the weaker will be the alignment strength  required to cause the reversal, and vice versa. Similar results are found for the dependence of the population reversal on the mean particle number density, as in a more dilute (denser) system the reversal is realized for a stronger (weaker) alignment strength.   

We study the implications of the aforementioned effect for the chirality-induced current at and swim pressure on the inner and outer boundaries of the annulus.  As generally expected, particle chirality leads to suppression of activity-induced effects such as boundary-accumulation of active particles and, as such, chiral active particles require a larger alignment strength to establish population reversal relative to nonchiral active particles. A remarkable effect emerges in the case of near-boundary chirality-induced currents. While for alignment strengths sufficiently far from the onset of population reversal, we find rotational current of a single well-defined sign  (clockwise or counterclockwise)   forming  near each of the inner and outer boundaries (albeit with opposing signs on the inner relative to the outer boundary), the situation turns out to be more complex near the onset of the population reversal; hence,  the currents near the two boundaries can take similar signs and thus rotate in the same direction,  and multiple `layers' of rotating currents can be seen, especially for weak chirality strengths. 

As the surface-population reversal is directly caused by the presence of interparticle alignment, it may be tempting to compare it with the bulk flocking transition in Vicsek-type models \cite{Vicsek}. Our results indicate that the confinement and boundary curvature play key roles in regulating the population reversal, making it possible to compare it also with surface-induced transitions, e.g., in wetting  \cite{Sepu,Joanny} and capillary   \cite{Wysocki,Kne} systems, and in counterion condensation phenomena \cite{Naji,Naji_2}. 

Our model treats the problem at hand on the level of a minimal model of active Brownian particles \cite{Marchetti_2}.  Hence, it neglects several other important factors that could be considered for a more comprehensive analysis in the future. These include the roles of hydrodynamic interactions between particles and between particles and the boundaries (see, e.g., Refs. \cite{Li_3,Elgeti}). In these contexts, an interesting problem would be that of the so-called pusher and puller microswimmers with dipolar flow fields (see, e.g., Refs. \cite{Ishikawa,Lauga,Pooley}) and how the ensuing active stress due to these markedly different types of active particles might influence the behaviors predicted with the current  setting, especially at elevated area fractions.  Since the far-field hydrodynamic interactions may be  screened by  confinement effects  \cite{Delfau}, the near-field hydrodynamics will be of particular interest in the analysis of active particle distributions within an annulus. The interparticle alignment due to such hydrodynamic effects may thus compete or cooperate with the dot-product alignment model considered here, paving the way for more intriguing possible scenarios. When the rodlike nature of active particles is accounted for,  particle interactions  with the boundaries will not be torque-free anymore and the swim pressure can vary depending on the type of surface potentials  \cite{Solon_2,Wang}. Active rods in circular geometries   \cite{Van,Vladescu} can also be subjected to imposed shear flows, constituting another potential direction of research that can be explored in the future. 

 \section{Conflicts of interest}

There are no conflicts of interest to declare. 

\section{Acknowledgements}
Z.F. thanks A. Partovifard and M. R. Shabanniya for useful discussions and comments. A.N. acknowledges partial support from the Associateship Scheme of The Abdus Salam International Centre for Theoretical Physics (Trieste, Italy). We thank the High Performance Computing Center of the Institute for Research in Fundamental Sciences (IPM)  for computational resources.

\section{Author contributions}
Z.F. performed the theoretical derivations and numerical coding, generated the output data and produced the figures. Both authors analyzed the results, contributed to the discussions and wrote the manuscript. A.N. conceived the study and supervised the research. 

\appendix

\section{Choice of parameter values}
\label{app:parameters}

We fix the interparticle and particle-wall steric interaction strengths at representative values of $\tilde{U}_1=10$ and $\tilde k_1=\tilde k_2=10$ and vary  other system parameters with a range representative values (i.e., ${\rm Pe}=1-10$,  $\tilde{R}_1,\tilde{R}_2=10-100$, $U_0=0-100$, $\tilde{\rho}_0=0.01-0.1$, $\Gamma=0-6$) to explore different regions of the parameter space. 
The main advantage of using dimensionless values is that they can be mapped to a wider range of actual parameter values relevant to realistic cases of synthetic and biological active particles  \cite{Bechinger,Berg,Jiang,Bricard,Darnton,Howse}. For instance, choosing $a\simeq 1\,{\rm\mu m}$ and $D_{\mathrm{r}}$ in the range $D_{\mathrm{r}}\simeq 0.1\,{\rm s^{-1}}$ (thermal diffusivity) to $1\,{\rm s^{-1}}$ (active tumbling), the range of P{\'e}clet numbers used here to discuss the representative behavior of the system can be mapped to self-propulsion speeds of up to  $v\simeq 10{\rm\mu m/s}$. 
Artificial active particles can take a wide range of chirality strengths as well, with examples furnished by curved self-propelled  rods ($\vert\Gamma\vert\simeq1-5$) \cite{Takagi,Takagi_2}, self-assembled rotors ($\vert\Gamma\vert\simeq1-13$) \cite{Wykes}, and Janus doublets ($\vert\Gamma\vert\simeq8-26$) \cite{Ebbens}. 
Our investigated range of rescaled mean densities ($\tilde \rho_0=0.01-0.1$) can also be compared with those   in  Refs. \cite{Vladescu,Van}. 

\section{Analytic expressions for swim pressure}\label{app1}

Here, we derive useful analytical expressions for the swim pressure for the system of confined interacting particles described  in the text.  Swim pressure on each of the inner/outer circular boundaries can be calculated by integrating the force exerted per unit length of that boundary, which was achieved by numerically calculating the PDF as discussed in the text; see Section  \ref{pressure}. The said analytical expression can be established as follows. 

Starting from the Smoluchowski equation \eqref{eq:SE} and writing it in polar coordinates ($r,\psi$), we  have
\begin{widetext}
\begin{equation}\label{Smol}
\begin{split}
\partial_t{\cal P}=-\frac{1}{r}\Bigg\{&\partial_r\Bigg(r\Bigg[\left(v\cos\psi-\mu_{\mathrm{t}}\partial_r{\mathcal V}-\mu_{\mathrm{t}}\partial_r\overline{{\mathcal U}}\right){\cal P}-D_{\mathrm{t}}\partial_r{\cal P}\Bigg]\Bigg)\\
&+\partial_\psi\left[\left(-v\sin\psi+r\omega-\frac{1}{r}\left(\mu_{\mathrm{t}}+r^2\mu_{\mathrm{r}}\right)\partial_\psi\overline{{\mathcal U}}\right){\cal P}-\frac{1}{r}\left(D_{\mathrm{t}}+r^2D_{\mathrm{r}}\right)\partial_\psi{\cal P}\right]\Bigg\},
\end{split}
\end{equation}
\end{widetext}
The steady-state local density of particles follows as the zeroth angular moment of the PDF with $\partial_t\rho=0$. Hence,   integrating the above equation over the angular coordinate and substituting the definition of the swim pressures from Eq. \eqref{pre}, we find these latter quantities in terms of the mean particle density and the local orientational order parameter (first moment) defined through Eq \eqref{eq:m1}. Thus, for the inner boundary (with the same procedure being applicable to $P_2$, not to be repeated here), we have 
\begin{equation}\label{pre2}
P_1=\frac{\rho_0}{\mu_{\mathrm{t}}}\left[D_{\mathrm{t}}\left(1+\frac{\rho_0U_1}{2}\right)\right]+\frac{v}{\mu_{\mathrm{t}}}\int_{\Lambda}^0\rmd r\,m_1(r).
\end{equation}
Multiplying both sides of Eq. \eqref{Smol} by  $\cos\psi$ and then integrating over the angular coordinate gives the steady-state equation, governing the first moment of equation (\ref{Smol}) in the steady state and in the regime of  sufficiently large radii of curvature as
\begin{widetext}
\begin{equation}
\begin{split}
D_{\mathrm{r}}m_1\simeq &\frac{1}{r}\left(-vm_2+\mu_{\mathrm{t}} m_1 \partial_ r{\mathcal V}+D_{\mathrm{t}}\partial_rm_1+U_1D_{\mathrm{t}}m_1\partial_r\rho\right)-\partial_r\left(\frac{v}{2}(\rho+m_2)-\mu_{\mathrm{t}} m_1 \partial_ r{\mathcal V}-D_{\mathrm{t}}\partial_rm_1-U_1D_{\mathrm{t}}m_1\partial_r\rho\right)\\
&-\omega\int\rmd \psi\,\sin\psi{\cal P}_0(r,\psi)-U_0D_{\mathrm{r}}\int \rmd r'\int\rmd \psi'\,r'{\cal P}_0(r',\psi')\int\rmd \psi\sin\psi\sin(\psi'-\psi){\cal P}_0(r,\psi),
\end{split}
\end{equation}
\end{widetext}
where $m_2(r)$ is the second angular moment of the PDF. Using the above relation in Eq. (\ref{pre2}), we  arrive at the approximate decomposition for the swim pressure as $P_1\simeq P_f+P_{ch}+P_{ex}+P_{int}$, with the definitions 
\begin{widetext}
\begin{eqnarray}
\label{eq:Pf}
&&P_f=\frac{\rho_0}{\mu_{\mathrm{t}}}\left(D_{\mathrm{t}}+\frac{U_1\rho_0D_{\mathrm{t}}}{2}+\frac{v^2}{2 D_{\mathrm{r}}}\right), \\
&&P_{ch}=-\frac{v\omega}{\mu_{\mathrm{t}}D_{\mathrm{r}}}\int_{-\pi}^{\pi}\int_{\Lambda}^0\rmd \psi\rmd r\sin\psi{\cal P}_0(r,\psi),\\
&&P_{ex}=\frac{v}{\mu_{\mathrm{t}}D_{\mathrm{r}}}\int_{\Lambda}^0\rmd r\,\frac{1}{r}\left(-vm_2+\mu_{\mathrm{t}} m_1 \partial_ r{\mathcal V}+D_{\mathrm{t}}\partial_rm_1+U_1D_{\mathrm{t}}m_1\partial_r\rho\right),\\
\label{P_int}
&&P_{int}=-\frac{vU_0}{\mu_{\mathrm{t}}}\int_{\Lambda}^0 \rmd r'\int_{-\pi}^{\pi}\rmd \psi'\,r'{\cal P}_0(r',\psi')\int_{\Lambda}^0 \rmd r\int_{-\pi}^{\pi}\rmd \psi\sin\psi\sin(\psi'-\psi){\cal P}_0(r,\psi).
\end{eqnarray}
\end{widetext}
 The term  $P_{f}$ gives the standard swim pressure of nonchiral active particles on a {\em flat} boundary and $P_{ex}$ is the excess pressure due to the curvature of the circular boundaries \cite{Jamali}. The first term in the expression for $P_f$ on the r.h.s. of Eq. \eqref{eq:Pf} is the ideal gas pressure, the second term is the second virial term due to steric interactions between the particles that scale with $U_1$, while the last term is the corresponding nonequilibrium swim pressure. $P_{ex}$ also involves an additional term explicitly dependent on the  steric particle interactions, but   neither of the three contributions $P_{f}$, $P_{ch}$ and $P_{ex}$ vary explicitly with  the alignment interactions, which enter implicitly in nonideal terms through the solution of the PDF. It is only the last term $P_{int}$ that explicitly scales with the alignment interaction strength, $U_0$. $P_{ch}$ is also the only term that  explicitly scales with the chirality strength.  Our numerical results show that $P_{int}$ is positive on the inner boundary and negative on the outer one. This is consistent with our findings in Fig. \ref{fig6}b, indicating that, for nonzero $U_0$, the total swim pressure on the inner (outer) boundary $P_{1}$ ($P_2$) is always larger (smaller) than its value in the zero-alignment case.  

\bibliography{references}

\end{document}